# Privacy-Preserving Decentralized Energy Management for Networked Microgrids via Objective-Based ADMM

Jesus Silva-Rodriguez, *Student Member IEEE*, and Xingpeng Li, *Senior Member IEEE*

***Abstract*****-- This paper proposes a decentralized energy management (DEM) strategy for a network of local microgrids, providing economically balanced energy schedules for all participating microgrids. The proposed DEM strategy can preserve the privacy of each microgrid by only requiring them to share network power exchange information. The proposed DEM strategy enhances the traditional alternating direction method of multipliers (ADMM) formulation for networks of microgrids by examining the global objective value as well as the solution quality. A novel stopping decision combining these two metrics is proposed for the enhanced objective-based ADMM (OB-ADMM) method. This paper also presents a centralized energy management (CEM) model as a benchmark, and a post-processing proportional exchange algorithm (PEA) to balance the economic benefit of each microgrid. The resulting proposed OB-ADMM model combined with the PEA delivers the final high-fidelity optimal solution for multiple microgrids in a grid-connected collaborative power exchange network. Moreover, the proposed decentralized operational strategy preserves the economic and privacy interests of individual microgrid participants. Different network cases are simulated to test the algorithm's performance, and the results validate the aforementioned claims.**

## NOMENCLATURE

*Sets*

$M$: Local microgrids in the network.
$G$: Generator units.
$ES$: Energy storage units.
$T$: Time intervals.

*Indices*

$m$: Local microgrid index.
$n$: Neighbor microgrid index.
$g$: Generator unit index.
$b$: Energy storage unit index
$t$: Time interval index.
$k$: Iteration index.

*Parameters*

$P^G_{min}$: Minimum generator output.
$P^G_{max}$: Maximum generator output
$SU^G$: Start-up cost of a generator.
$NL^G$: No-load cost of a generator.
$C^G$: Operating cost of a generator.
$P^{ES}_{lim}$: Power rating of an energy storage unit.
$\eta^{ESd}$: Discharging efficiency of an energy storage unit.
$EL^{ES}_{min}$: Minimum charge level of an energy storage unit.
$EL^{ES}_{max}$: Maximum charge level of an energy storage unit.
$P^L$: Power load demand of a microgrid.
$P^{SP}$: Solar power output of a microgrid.
$P^{WP}$: Wind power output of a microgrid.
$C^{grid+}$: Grid power import price.
$C^{grid-}$: Grid power export price.
$C^{Np}$: Price of power exchanged among microgrids.
$P^E_{lim}$: Microgrid tie-line limit to central node.
$\rho$: ADMM penalty parameter.
$\beta$: Average objective value rate of change threshold.
$k_s$: Algorithm iterations offset.

*Variables*

$P^G$: Generator power output.
$u^G$: Generator binary status indicator.
$v^G$: Generator binary start-up indicator.
$P^{ESc}$: Energy storage unit power input.
$P^{ESd}$: Energy storage unit power output.
$e^{ESc}$: Energy storage unit binary charging status indicator.
$e^{ESd}$: Energy storage unit binary discharging status indicator.
$EL^{ES}$: Energy storage unit charge level.
$P^{grid+}$: Power import from main grid.
$P^{grid-}$: Power export from main grid.
$P^{N+}$: Power import into one microgrid from another.
$P^{N-}$: Power export out of one microgrid to another.
$p^{N+}$: Microgrid binary power import indicator.
$p^{N-}$: Microgrid binary power export indicator.
$y^L$: ADMM Lagrange multiplier
$r^L$: Primal residual.
$s^L$: Dual residual.

## I. INTRODUCTION

Microgrids are small community-level power systems that feature various different distributed energy resources (DER) and loads and can operate independently [1]. Moreover, microgrids can operate connected or isolated from the main grid in the event that main grid contingencies occur. Therefore, the operation strategies that microgrids can implement further increase reliability and resilience of power distribution systems, in addition to benefits regarding economic, security, and sustainability aspects of the power distribution operation [2]-[3].

A network of microgrids can also increase the economic benefits provided by single microgrid operations. A network comprising adjacent nearby microgrids enables mutual power sharing and thus, it collectively increases reliability and reduces operational costs of the microgrids participating in the network [4]. These benefits have extensively been proven in the literature through the study on different microgrid network configurations and control schemes.

Jesus Silva-Rodriguez and Xingpeng Li are with the Department of Electrical and Computer Engineering, University of Houston, Houston, TX, 77204, USA (e-mail: jasilvarodriguez@uh.edu; xingpeng.li@asu.edu).

A cluster of microgrids interconnected in a network configuration may operate based on two different control models: a centralized or a decentralized operation. A centralized model manages and optimizes all microgrids in the network through a single control system, guaranteeing the optimal operation of all microgrids in a network through an efficient energy management due to its unlimited network information access. However, critical issues may arise due to its higher computational load and impact on operability when the central control system fails, in addition to information privacy concerns. In a decentralized model, on the other hand, each microgrid carries out its own energy management separately, and only communicate minimal information to the network while engaging in power sharing to achieve global network optimization in a distributed manner. In this way, a single component failure does not compromise all microgrids in the network [4]-[5].

Centralized energy management (CEM) of microgrid networks may be developed in ways that some of these issues are resolved directly in its formulation. A security-oriented approach to the optimal coupling of multiple microgrids is presented in [6], implementing an optimization approach that allows microgrids experiencing issues to recover via external support from neighboring microgrids. A model for a network of micro water-energy nexus systems is presented in [7], featuring a centralized network formulation which balances network exchanges to provide an equal economic benefit for all participants while maintaining the global optimal operation. An optimal procedure for the economic schedule of a network of microgrids is implemented in [8], using a distributed model predictive control algorithm that improves the cost function of each microgrid with the network operation. However, the privacy concerns are not yet resolved, and there is also a need to consider each microgrid as an individual entity willing to participate in the network only when improved economic benefits are obtained than acting as a single system [8]. As a result, decentralized network formulations are preferred for practical applications.

The alternating direction method of multipliers (ADMM) is an algorithm well suited for distributed convex optimization and is often applied to solve problems in which the optimization can be carried out locally, and then coordinated globally via specific constraints that are relaxed to find a solution [9]. A distributed control framework for a network of microgrids is proposed in [10], featuring a hierarchical two-level control approach. The lower-level tracks the optimal generation setting of each generator in each microgrid internally while the upper-level solves the optimal dispatch of all microgrids by decomposing the network to the microgrid level and cooperatively optimizing it by implementing ADMM.

ADMM has also been used to solve large scale problems such as the optimal power flow (OPF) problem in a distributed manner. Distributed algorithms based on ADMM have been implemented to formulate a decentralized approach to solving the OPF problem in [11]-[12], demonstrating the feasibility of solving OPF in a distributed manner as opposed to solving it as part of a large, centralized power system. Moreover, large-scale power systems can be optimized in a distributed manner by solving the AC OPF problem with a geographical decomposition of the network using ADMM along with a partitioning technique, achieving a solution very close to the local optimum efficiently [13].

Given the nature of the network of microgrids optimization problem, ADMM can also be used to implement a decentralized approach to solving this problem. In literature, this strategy has already been implemented using ADMM in different way. In [14]-[15], the distributed energy management method is developed based on a convex optimization problem for a multi-microgrid system implementing ADMM to decentralize the model, and reaching a system-wide optimum by implementing a duplication of network global variables, setting them equal to each other and using these equalities as the constraints relaxed to decentralize the model. Similarly, [16] presents a distributed power management framework based on ADMM for multi-microgrid networks in which the communication among microgrids is limited to only their power exchanges information. This approach is viable for complicated networked microgrid systems as well as distribution networks, however, the model ends up requiring essentially three different optimization problems as it first optimizes the energy management schedule of each microgrid, and then the local and global variables separately within its ADMM iterations.

An online energy management based on ADMM is also presented in [17], exploring the potential for the use of regret minimization for an energy management algorithm for networks of microgrids, decoupling the power flows and voltage levels of every microgrid in the network by duplicating these variables for each microgrid individual optimization. Furthermore, [17] implements ADMM to solve each time interval executing a single iteration of ADMM per interval, with regret minimization guaranteeing convergence as the algorithm iterates progress to the last time interval. This approach is mainly favorable when considering uncertainty, which is outside the scope of this paper.

A very important control aspect of the ADMM algorithm is its convergence, since this is what will determine whether the algorithm has arrived at the final solution. One advantage of ADMM is that it only has one control parameter, known as the *penalty*, which has direct impact on algorithm convergence [18]. This penalty term is defined according to the primal and dual residual terms of the algorithm, which are defined internally for ADMM based on how well the algorithm is meeting the relaxed constraint thus far [9], [18]. The ADMM applications of the literature implement an adaptive varying penalty strategy outlined in [9] which is able to speed up convergence in some cases. However, this strategy has been proven to present a serious flaw that arises primarily when attempting to implement a scaled problem due to the different scaling properties of the defined primal and dual residual terms. Moreover, no known relationship between the minimization of the primal and dual residuals and the actual solution suboptimality has been found, as this approach only guarantees feasibility of the problem by ensuring the relaxed constraint is met [19]. As a result, strategies that take solution optimality into account in addition to feasibility should be considered.

Therefore, after careful consideration of the current ADMM implementation methods in decentralized and distributed system optimization, this paper proposes a novel objective-based ADMM (OB-ADMM) decentralized energy management (DEM) approach in which both solution quality and optimality are combined to determine system-wide convergence to a guaranteed optimal solution. Moreover, considering that each microgrid in the network represents an independent entity, the algorithm is formulated in a way that only key power exchange information between microgrid pairs is communicated in the network, in addition to the implementation of a post processing algorithm that ensures an economically fair proportional exchange among the local microgrids and the main grid.

The rest of the paper is organized as follows. A base case CEM optimization model is formulated in Section II where a proportional exchange algorithm (PEA) that ensures an equally fair economic benefit for all microgrids is also introduced. Section III presents the decentralization of the model presented in Section II via ADMM, as well as proposed OB-ADMM algorithm for a network of microgrids. Section IV presents case studies used to test the algorithm for different scenarios and their corresponding results compared against the results obtained from centralized network formulations for the same test cases. Finally, Section V concludes the paper.

## II. Network Optimization Model

### A. Single Microgrid Energy Management

The energy management operation for a single microgrid is modeled as an optimization problem with the objective of minimizing the total cost associated with managing all energy sources and loads within the microgrid while satisfying system technical constraints [20].

A microgrid is typically composed of controllable generators and uncontrollable renewable energy sources such as solar and wind power, in addition to battery energy storage systems that further increase the microgrid system reliability and economic advantages. For grid-connected microgrids, the power import/export capabilities with the main grid also need to be included in the energy management strategy. The objective of the energy management problem will optimize the microgrid operation to fully meet its respective power demand, and as such the objective function of a single microgrid *m* is formulated in (1). This objective function involves all the control variables of the diesel generators and the imports and exports with the main grid, each with their respective costs.

$$f_{cost,m} = \sum_{t\in T}\{\sum_{g\in G}\left[SU^G_{m,g}v^G_{m,g,t} + \Delta t \cdot \left(NL^G_{m,g}u^G_{m,g,t} + C^G_{m,g}P^G_{m,g,t}\right)\right] + C^{grid+}_t P^{grid+}_{m,t} - C^{grid-}_t P^{grid-}_{m,t}\} \quad (1)$$

Constraints defining the operation limits of each microgrid component are also formulated. These ensure the microgrid is operating within appropriate nominal levels. For the generators, the power output of each unit is regulated by (2) and their status is controlled with (3). The input and output of the battery energy storage units are regulated by (4) and (5), respectively, and (6) ensures each unit is only either charging, discharging or idle at each time interval. Moreover, (7) and (8) calculate and limit the energy level of the energy storage units, respectively.

$$P^G_{min,m,g}u^G_{m,g,t} \le P^G_{m,g,t} \le P^G_{max,m,g}u^G_{m,g,t}, (\forall m \in M, g \in G, t \in T) \quad (2)$$

$$v^G_{m,g,t} \ge u^G_{m,g,t} - u^G_{m,g,t-1}, (\forall m \in M, g \in G, t \in T) \quad (3)$$

$$0 \le P^{ESc}_{m,b,t} \le P^{ES}_{lim,m,b}e^{ESc}_{m,b,t}, (\forall m \in M, b \in ES, t \in T) \quad (4)$$

$$0 \le P^{ESd}_{m,b,t} \le P^{ES}_{lim,m,b}e^{ESd}_{m,b,t}, (\forall m \in M, b \in ES, t \in T) \quad (5)$$

$$e^{ESc}_{m,b,t} + e^{ESd}_{m,b,t} \le 1, (\forall m \in M, b \in ES, t \in T) \quad (6)$$

$$EL^{ES}_{m,b,t} = EL^{ES}_{m,b,t-1} + \Delta t \cdot \left(\eta^{ESc}_{m,b}P^{ESc}_{m,b,t} - P^{ESd}_{m,b,t}/\eta^{ESd}_{m,b}\right), (\forall m \in M, b \in ES, t \in T) \quad (7)$$

$$EL^{ES}_{min,m,b} \le EL^{ES}_{m,b,t} \le EL^{ES}_{max,m,b}, (\forall m \in M, b \in ES, t \in T) \quad (8)$$

Additionally, following the key requirement of energy management systems to fully meet demand, an equality constraint formulated in (9) ensures that the combined output of the microgrid's energy sources fully meets the "net load," which is defined as the difference between load and renewable generation from the solar and wind resources in (10).

$$\sum_{g\in G}P^G_{m,g,t} + \sum_{b\in ES}\left[P^{ESd}_{m,b,t} - P^{ESc}_{m,b,t}\right] + P^{grid+}_{m,t} - P^{grid-}_{m,t} = P^{net}_{m,t}, (\forall m \in M, t \in T) \quad (9)$$

$$P^{net}_{m,t} = P^L_{m,t} - P^{SP}_{m,t} - P^{WP}_{m,t} \quad (10)$$

### B. Centralized Model Formulation

The CEM formulation for a group of coordinated adjacent microgrids is illustrated in Fig. 1. In this network configuration, all microgrids are capable of exchanging power with all other microgrids in the network as well as with the main grid since all are connected through a central node. As a result, a CEM global optimization model for a network of microgrids must incorporate power imports and exports to and from other microgrids in the local optimization of each microgrid [5].

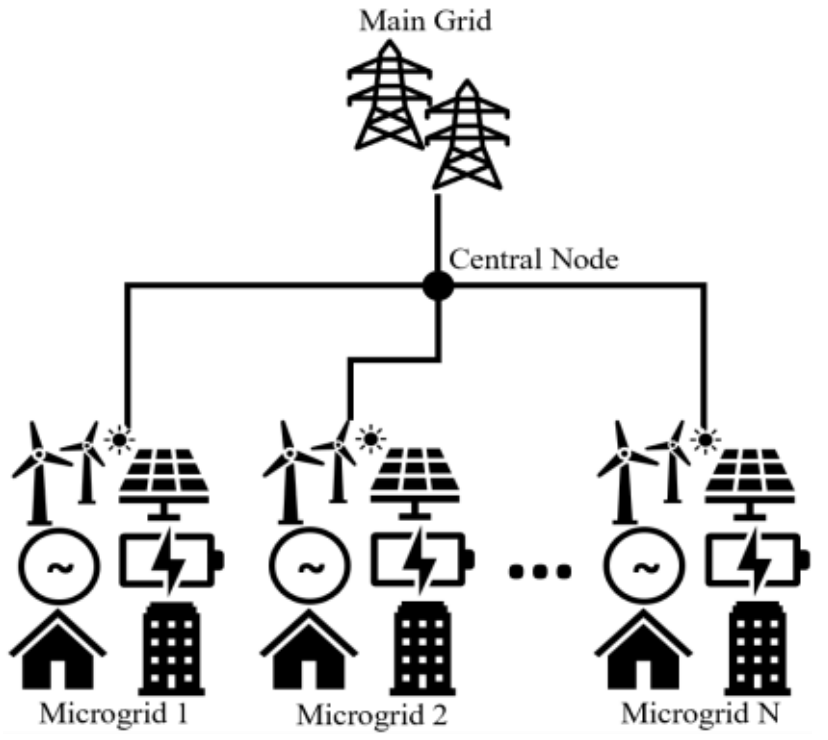

Fig. 1. Central node topology for a network of microgrids.

Parting from the energy management formulation for a single microgrid, the objective function of the CEM model is defined as shown in (11) and (12), including the power exchange with other microgrids. Moreover, the power balance of each microgrid from (9) is also updated to include the network exchanges and is defined in (13).

$$minimize \sum_{m\in M} f_{cost,m} \quad (11)$$

$$f_{cost,m} = \sum_{t\in T}\{\sum_{g\in G}\left[SU^G_{m,g}v^G_{m,g,t} + \Delta t \cdot \left(NL^G_{m,g}u^G_{m,g,t} + C^G_{m,g}P^G_{m,g,t}\right)\right] + C^{grid+}_t P^{grid+}_{m,t} - C^{grid-}_t P^{grid-}_{m,t} + C^{Np}_t \sum_{n\in M, n\ne m}\left(P^{N+}_{m,n,t} - P^{N-}_{m,n,t}\right)\} \quad (12)$$

$$\sum_{g\in G}P^G_{m,g,t} + \sum_{b\in ES}\left[P^{ESd}_{m,b,t} - P^{ESc}_{m,b,t}\right] + P^{grid+}_{m,t} - P^{grid-}_{m,t} + \sum_{n\in M, n\ne m}\left(P^{N+}_{m,n,t} - P^{N-}_{m,n,t}\right) = P^L_{m,t} - P^{SP}_{m,t} - P^{WP}_{m,t}, (\forall m \in M, t \in T) \quad (13)$$

Constraints reflecting the network interconnectivity and capacity limits must also be formulated in order to establish operation limits regarding the power flows through the central node in and out of the microgrids, and to and from the main grid. These are defined in (14)-(21).

$$P_{m,t}^{E} = P_{m,t}^{grid+} - P_{m,t}^{grid-} + \sum_{n\in M, n\neq m}\left(P_{m,n,t}^{N+} - P_{m,n,t}^{N-}\right), (\forall m \in M, t \in T) \quad (14)$$

$$-P_{lim,m}^{E} \leq P_{m,t}^{E} \leq P_{lim,m}^{E}, (\forall m \in M, t \in T) \quad (15)$$

$$0 \leq P_{m,n,t}^{N+} \leq P_{lim,m}^{E} p_{m,t}^{N+}, (\forall m, n \in M, n \neq m, t \in T) \quad (16)$$

$$0 \leq P_{m,n,t}^{N-} \leq P_{lim,m}^{E} p_{m,t}^{N-}, (\forall m, n \in M, n \neq m, t \in T) \quad (17)$$

$$0 \leq P_{m,t}^{grid+} \leq P_{lim,m}^{E} p_{m,t}^{N+}, (\forall m \in M, t \in T) \quad (18)$$

$$0 \leq P_{m,t}^{grid-} \leq P_{lim,m}^{E} p_{m,t}^{N-}, (\forall m \in M, t \in T) \quad (19)$$

$$p_{m,t}^{N+} + p_{m,t}^{N-} \leq 1, (\forall m \in M, t \in T) \quad (20)$$

$$P_{m,n,t}^{N+} = P_{n,m,t}^{N-}, (\forall m, n \in M, n \neq m, t \in T) \quad (21)$$

The total power imported into each microgrid is defined by (14), and the power transfer limit for each microgrid's tie-line to the central node is enforced in (15). The power exchanges from one microgrid to the other are regulated by (16) and (17), representing power import and export with the other microgrids, respectively. Similarly, the power coming from and to the main grid for each microgrid is limited by the same tie-line and thus, this is reflected by (18) and (19), respectively. The import and export status of each microgrid is controlled with (20) to prevent the erroneous status of simultaneous import and export from the optimization result. And finally, the nodal power balance of the central node is described by (21), ensuring that the import of power to one microgrid *m* from another microgrid *n* is equal to the export of microgrid *n* to microgrid *m*. This constraint (21) is considered as a "global constraint" since in the reference of each individual microgrid it would require a power exchange variable from another microgrid, requiring them to be on constant communication.

Executing this mixed-integer linear programming (MILP) problem minimizing the cost of all microgrids at once would correspond to solving the CEM model with the global energy management performed by the same system, requiring knowledge of all the variables of the participating microgrids. The result would correspond to an optimal solution that minimizes the cost of managing all microgrids combined.

### C. Proportional Exchange Algorithm

If scenarios in which the costs of trading with the main grid are less beneficial than the costs of trading within the network are considered, such as the case presented in [10], then the optimal solution that results from the CEM optimization model may not necessarily represent the solution that guarantees all microgrids are obtaining an equally fair economic benefit from their network participation. Therefore, a post-processing step is needed in which the solution is modified to balance the power exchanges among all microgrids and the main grid. This step should guarantee that each microgrid is enjoying an economic benefit derived from network power exchanges proportional to their local import and export requirements.

The PEA detailed in Algorithm 1 resolves this economic benefit discrepancy and is derived based on the algorithm utilized in [7] for a centralized network of micro water-energy nexus systems, with major modifications to account for the individual import/export variable definitions for each microgrid required for ADMM implementation.

When the cost of buying and selling power between the microgrids in the network is set to a price in between the purchase and selling prices of the main grid [10], the PEA adjusts the current solution from the optimization based on a proportion of their power import and export needs and the total power available for import and export from other microgrids in the network. This will ensure that not a single microgrid will receive all the available power for import in the network leaving the others to trade with the main grid at less favorable prices.

Algorithm 1: PEA for power exchange in networks of microgrids.

Solve the network of microgrids MILP problem and obtain power exchanges $P_{m,n,t}^{N+}$ and $P_{m,n,t}^{N-}$, and the net exchanges of each microgrid $P_{m,t}^{E}$.<br>
Allocate space for new variables $P_{m,t}^{E+}$ and $P_{m,t}^{E-}$<br>
**For** *t* in *T*<br>
  **For** *m* in *M*<br>
    **If** $p_{m,t}^{N+} = 1$<br>
      Set $P_{m,t}^{E+} = \left|P_{m,t}^{E}\right|$ and $P_{m,t}^{E-} = 0$<br>
    **Else**<br>
      Set $P_{m,t}^{E+} = 0$ and $P_{m,t}^{E-} = \left|P_{m,t}^{E}\right|$<br>
  **end For**<br>
  **For** *m* in *M*<br>
    **If** $p_{m,t}^{N+} = 1$<br>
      **If** $\sum_{m\in M} P_{m,t}^{E+} > \sum_{m\in M} P_{m,t}^{E-}$<br>
        **For** *n* in *M* ($m \neq n$)<br>
          $P_{m,n,t}^{N+} = \frac{P_{m,t}^{E+}}{\sum_{m\in M} P_{m,t}^{E+}} \cdot P_{n,t}^{E-}$<br>
        **end For**<br>
      **Else**<br>
        **For** *n* in *M* ($m \neq n$)<br>
          $P_{m,n,t}^{N+} = \frac{P_{m,t}^{E+}}{\sum_{m\in M} P_{m,t}^{E-}} \cdot P_{n,t}^{E-}$<br>
        **end For**<br>
      Set $P_{m,t}^{grid+} = P_{m,t}^{E+} - \sum_{n\in M, n\neq m} P_{m,n,t}^{N+}$ and $P_{m,t}^{grid-} = 0$<br>
    **Else**<br>
      **If** $\sum_{m\in M} P_{m,t}^{E+} < \sum_{m\in M} P_{m,t}^{E-}$<br>
        **For** *n* in *M* ($m \neq n$)<br>
          $P_{m,n,t}^{N-} = \frac{P_{m,t}^{E-}}{\sum_{m\in M} P_{m,t}^{E-}} \cdot P_{n,t}^{E+}$<br>
        **end For**<br>
      **Else**<br>
        **For** *n* in *M* ($m \neq n$)<br>
          $P_{m,n,t}^{N-} = \frac{P_{m,t}^{E-}}{\sum_{m\in M} P_{m,t}^{E+}} \cdot P_{n,t}^{E+}$<br>
        **end For**<br>
      Set $P_{m,t}^{grid-} = P_{m,t}^{E-} - \sum_{n\in M, n\neq m} P_{m,n,t}^{N-}$ and $P_{m,t}^{grid+} = 0$<br>
  **end For**<br>
**end For**

## III. DEM Strategy

### A. Alternating Direction Method of Multipliers Theory

ADMM is often applied to solve problems where the optimization can be carried out locally, and then coordinated globally via network constraints. For problems of the form shown in (22), this is done by implementing a network decomposition that separates the objective function into several subproblems for each local function and carrying out its optimization separately using the augmented Lagrangian (23), which relaxes the global constraints and adds them to the objective function with assigned penalties. Once all local subproblems have been executed, the results of the global variables are communicated to the other subsystems to be used as parameters for the next ADMM iteration [9].

The ADMM iterations must be formulated in an alternating and sequential manner [9], [21]. This means that within the same iteration, the individual systems must be optimized in a predefined order, with each utilizing the most up to date information from the previously optimized system. Following this approach, the ADMM iterations for a network of three systems would then be formulated in the form shown in (24) [21]. For this formulation, the algorithm will require *clock synchronization at the network level* to correctly run the optimization of each system in the network [12].

$$\begin{aligned} &minimize\ f(x) = \textstyle\sum_{i\in N} f_i(x_i) \\ &subject\ to\ \textstyle\sum_{i\in N} A_i x_i = b \end{aligned} \tag{22}$$

$$L(x,y) = \textstyle\sum_{i\in N} f_i(x) + \sum_{i\in N} y^T(A_i x_i - b) + \frac{\rho}{2}\|\sum_{i\in N}(A_i x_i - b)\|_2^2 \tag{23}$$

$$\begin{aligned} x_1^{k+1} &= argmin\ L(x_1, x_2^k, x_3^k, y^k) \\ x_2^{k+1} &= argmin\ L(x_1^{k+1}, x_2, x_3^k, y^k) \\ x_3^{k+1} &= argmin\ L(x_1^{k+1}, x_2^{k+1}, x_3, y^k) \\ y^{k+1} &= y^k + \rho(A_1 x_1^{k+1} + A_2 x_2^{k+1} + A_2 x_2^{k+1}) \end{aligned} \tag{24}$$

ADMM needs to be carried out in a sequential manner because the augmented Lagrangian of (23) is not conditionally independent, since the Lagrange multipliers are composed of all global variables. Thus, a simultaneous optimization of each system would give different, usually worse results than the optimization of all systems sequentially [21].

In addition to the variable updates of (24), the primal and dual residuals at every iteration should also be determined and are given by (25) and (26), respectively [10], [14].

$$r^{k+1} = \textstyle\sum_{i\in N}\left(A_i x_i^{k+1} - b\right) \tag{25}$$

$$s^{k+1} = \textstyle\sum_{i\in N}\left(A_i x_i^{k+1} - A_i x_i^k\right) \tag{26}$$

The primal and dual residuals help determine the convergence of ADMM, since ideally when these two metrics reach zero the algorithm should be at a feasible solution. Therefore, the primal and dual residuals can be used as stopping criteria. However, the relationship between these stopping criteria and the actual solution suboptimality is unknown, therefore convergence to zero of these two metrics does not necessarily imply convergence to the optimal solution [19], hence the need for additional stopping criteria.

### B. DEM Model Formulation via ADMM

The ADMM algorithm can effectively be used to decentralize the network optimization model presented in Section II. Since the objective function given by (11) represents the sum of *M* microgrid objective functions, it can be decomposed into individual functions optimized separately. Most of the model constraints can also be clearly decomposed into independent subproblems for each microgrid *m*. However, any constraints that once separated still require information on variables from other microgrids will need to be relaxed, with those variables treated as parameters that get updated through the iterations of ADMM. In this case, (21) corresponds to said constraint, therefore it is used to update the objective function of each microgrid into an augmented Lagrangian. Based on ADMM theory [9], [18], [21], the augmented Lagrangian then should take the form shown in (25) for this DEM problem.

$$L = \textstyle\sum_{m\in M} f_{cost,m} + \sum_{t\in T}\sum_{m\in M}\sum_{n\in M}^{n\neq m}\left[y_{m,n,t}^L\left(P_{m,n,t}^{N+} - P_{n,m,t}^{N-}\right) + \frac{\rho}{2}\left(P_{m,n,t}^{N+} - P_{n,m,t}^{N-}\right)^2\right] \tag{25}$$

Using (22) as the local objective function for each microgrid optimization, an initial ADMM algorithm for a network of microgrids is formulated, using a solution quality metric $\varepsilon$ based on the primal and dual residuals defined by (26) as the stopping criterion [19].

$$\varepsilon^k = \sqrt{\left\|r^{L^k}\right\|_2^2 + \left\|s^{L^k}\right\|_2^2} \tag{26}$$

The ADMM formulation detailed in Algorithm 2 is used as a reference to study its convergence behavior and determine an additional stopping criterion customized for efficiently solving this DEM problem that, along with the solution quality metric, will guarantee the algorithm converges towards the actual optimal solution of the problem.

Algorithm 2: Reference ADMM algorithm formulation

1. Initialize algorithm variables: $P_{m,n}^{N+^0} = P_{init,m,n}^{N+}$ ; $P_{m,n}^{N-^0} = P_{init,m,n}^{N-}$ ; $y_{m,n}^{L^0} = 0$ ; $r_{m,n}^{L^0} = 1$ ; $s_{m,n}^{L^0} = 1$ ; $\rho = \rho_0$; $k = 1$
2. **While** $\sqrt{\left\|r^{L^k}\right\|_2^2 + \left\|s^{L^k}\right\|_2^2} > \varepsilon_{th}$
3. **For** *m* in *M*
4. Solve individual microgrid optimization: $\left[P_{m,n}^{N+^{k+1}}, P_{m,n}^{N-^{k+1}}\right] = argmin_{P_{m,n}^{N+},P_{m,n}^{N-}}\left\{f_{cost,m} + \sum_{n\in M,n\neq m}\left[y_{m,n}^{L^k}\left(P_{m,n}^{N+} - P_{n,m}^{N-^{k+1}}\right) + y_{n,m}^{L^k}\left(P_{n,m}^{N+^{k+1}} - P_{m,n}^{N-}\right) + \frac{\rho}{2}\left(\left(P_{m,n}^{N+} - P_{n,m}^{N-^{k+1}}\right)^2 + \left(P_{n,m}^{N+^{k+1}} - P_{m,n}^{N-}\right)^2\right)\right]\right\}$
5. **end For**
6. **For** *m* in *M*
7. **For** *n* in *M* ($m \neq n$)
8. $y_{m,n}^{L^{k+1}} = y_{m,n}^{L^k} + \rho\left(P_{m,n}^{N+^{k+1}} - P_{n,m}^{N-^{k+1}}\right)$
9. $r_{m,n,t}^{L^{k+1}} = P_{m,n}^{N+^{k+1}} - P_{n,m}^{N-^{k+1}}$
10. $s_{m,n}^{L^{k+1}} = \left(P_{m,n}^{N+^{k+1}} - P_{n,m}^{N-^{k+1}}\right) - \left(P_{m,n}^{N+^k} - P_{n,m}^{N-^k}\right)$
11. **end For**
12. **end For**
13. $P_{m,n}^{N+^{k+2}} = P_{m,n}^{N+^{k+1}}$
14. $P_{m,n}^{N-^{k+2}} = P_{m,n}^{N-^{k+1}}$
15. $k = k + 1$
16. **end While**

The global variable duplication and allocation for two iterations later on steps 13 and 14 of Algorithm 2 are included so the sequential approach to ADMM can be achieved. In this way, step 4 is guaranteed to start with power exchange values of the previous iteration, and then the next microgrid receives the immediate exchange information from the previous microgrid to update the parameters for its own optimization.

### C. Proposed OB-ADMM Algorithm

The form of ADMM proposed in this paper is an "objective-based" approach: in addition to the solution quality metric $\varepsilon$, the global objective value of the network optimization obtained in every iteration is also considered. The DEM strategy consists of an analysis on the decrease on the rate of change of the objective value along with inspection of the solution quality by considering the metric $\varepsilon$. Once the average objective value's rate of change for a past number of iterations $k_s$ has decreased down to a threshold $\beta$, the algorithm considers the solution quality at the current iteration and stops the algorithm once it determines $\varepsilon$ is below the average solution quality metric in the past $k_s$ iterations. This strategy is appended to the reference ADMM form of Algorithm 2 and the final OB-ADMM algorithm is represented graphically in Fig. 2.

This OB-ADMM algorithm replaces the CEM optimization model from Section 2 and delivers a global optimal solution that again does not necessarily reflect the specific optimal solution that benefits all microgrids equally. Thus, the PEA needs to be implemented as a post process in the same way as for the CEM model. This will ensure the final solution ob-tained will always correspond to the one benefiting all participants fairly and should correspond to the exact same solution as the one obtained with the CEM optimization.

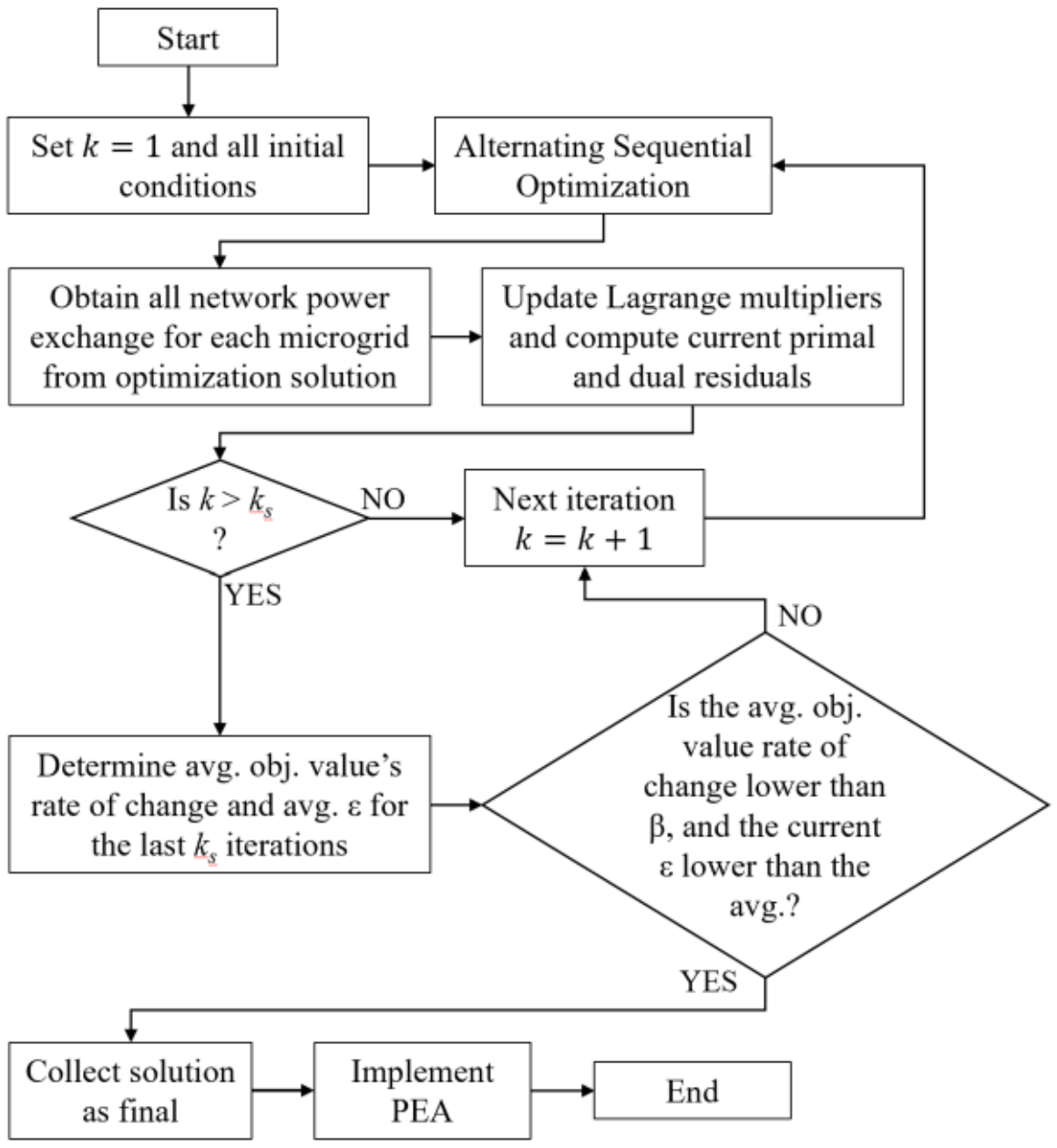

Fig. 2. Proposed OB-ADMM formulation for DEM model.

## IV. Case Studies

Three different test cases for networks of microgrids are designed and tested to validate the proposed DEM strategy and OB-ADMM algorithm. All relevant data and information about these cases is included in the Appendix.

### A. CEM Model

The CEM model formulation for a group of networked microgrids is utilized as a reference. Benchmark scenarios are derived for the DEM optimization via ADMM results to be compared against and determine whether this method approaches the optimal solution obtained by the CEM. Moreover, the CEM model makes use of the PEA to adjust the final solution to ensure a fair economic benefit for all participants. Therefore, the results presented in this subsection correspond to the power exchange distribution among microgrids for network cases 1, 2, and 3 from the Appendix, and are plotted in Figs. 3, 4 and 5, respectively. These power exchanges do not include main grid imports/exports, and only focus on the exchange interaction within the network. Moreover, the objective value results breakdown by microgrid as well as the total global objective values for each network case are summarized in Table I.

### B. ADMM Convergence and Performance

To test the performance and convergence of the reference form of ADMM given in Algorithm 2, network case 1 is used. Given that ADMM's only parameter with direct impact on convergence is the penalty $\rho$ [18], a sensitivity analysis of this parameter is carried out to see how the ADMM algorithm converges with different values of $\rho$ for this problem with this test case. Moreover, the results obtained with the CEM model for the same test case are used as reference to determine the quality of the solution obtained with the proposed OB-ADMM approach: how close it is to the optimal solution.

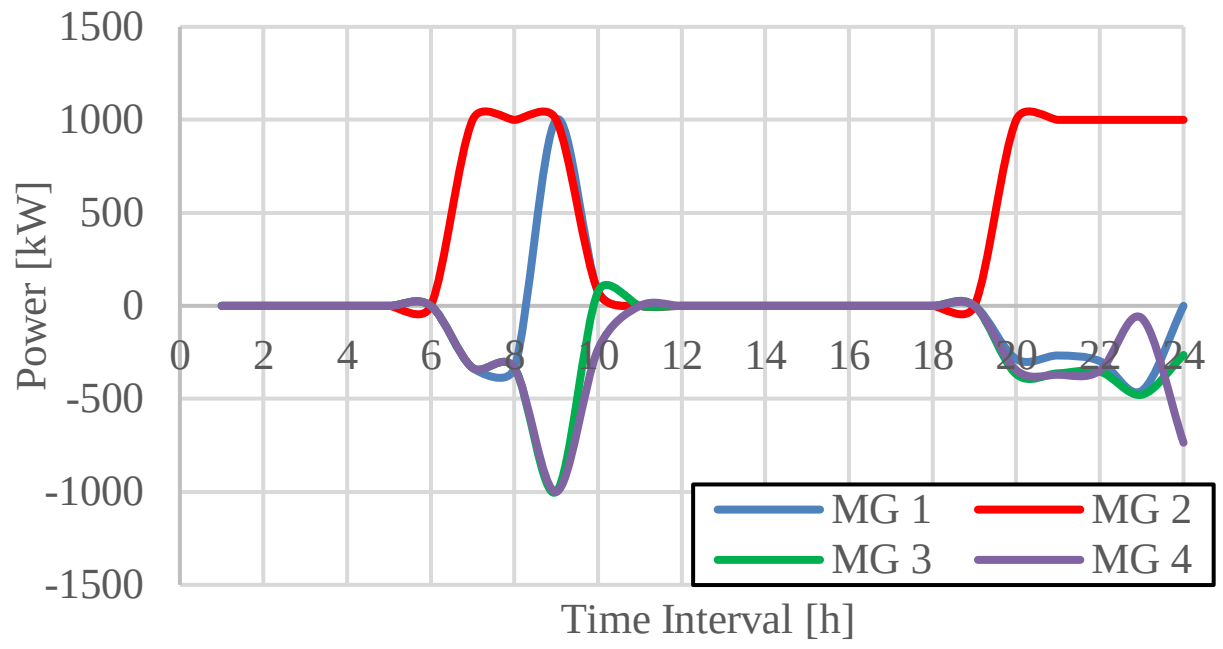

Fig. 3. Case 1 network power exchanges among microgrids.

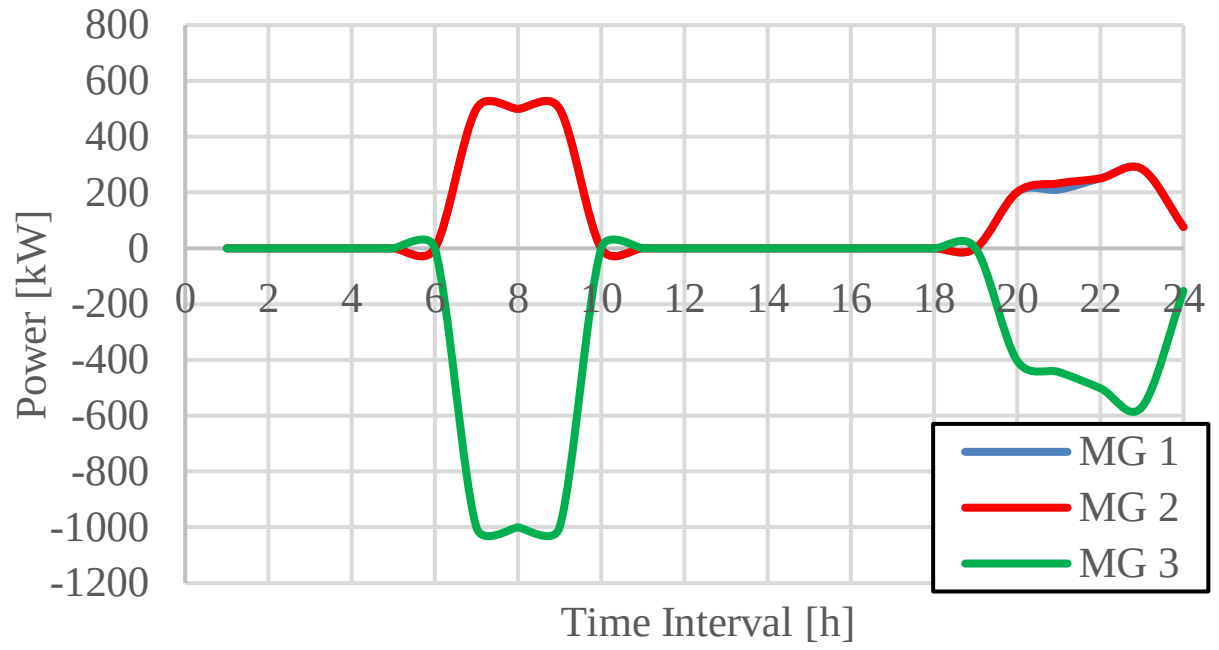

Fig. 4. Case 2 network power exchanges among microgrids.

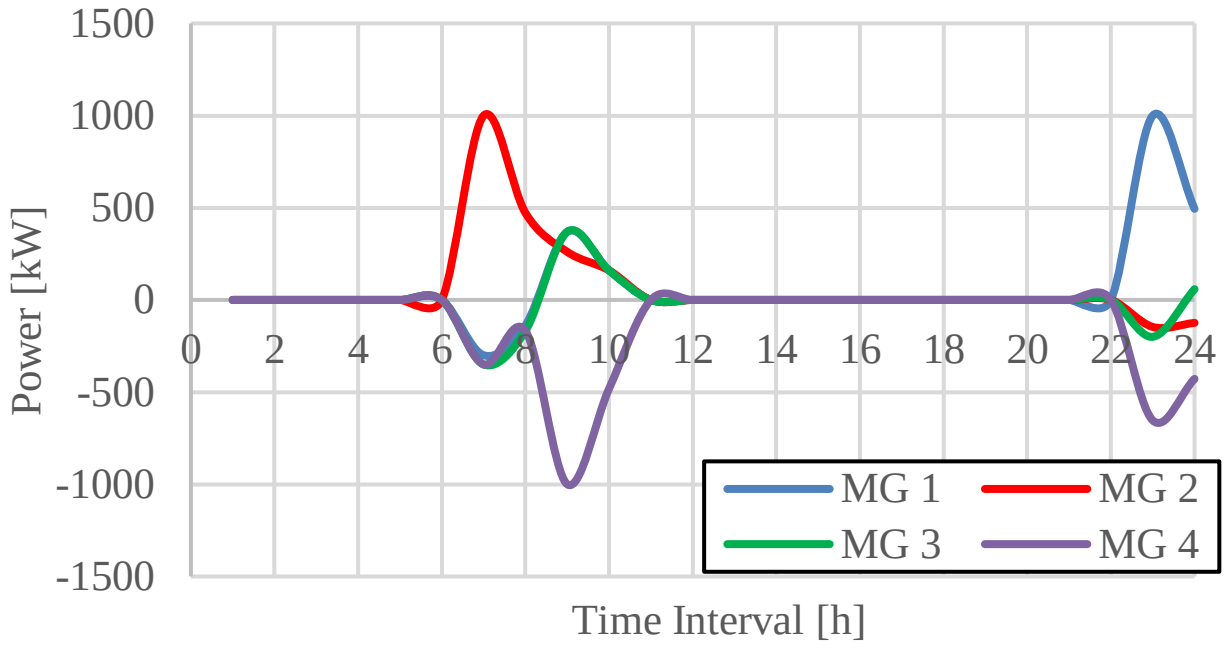

Fig. 5. Case 3 network power exchanges among microgrids.

Table I. Objective value results of centralized model.

| Microgrid | Case 1 | Case 2 | Case 3 |
|---|---|---|---|
| 1 | 241.95 | 1153.95 | 499.99 |
| 2 | 3258.83 | 3221.97 | 552.20 |
| 3 | 71.95 | 106.92 | 234.54 |
| 4 | 30.37 | N/A | -2.77 |
| TOTAL | 3603.10 | 4482.84 | 1283.96 |

Analysis of the solution quality metric $\varepsilon$ indicates that the algorithm reduces $\varepsilon$ faster with larger values of $\rho$. However, this fast $\varepsilon$ reduction does not guarantee convergence to the optimal value of the objective function. Table II shows how many iterations it takes to decrease $\varepsilon$ lower than 0.1 per penalty value, along with the ratio of the objective value obtained with the actual optimal value from the centralized model objective value results.

Looking at the final objective value of the solutions as shown in Table II, it is noticeable that convergence of $\varepsilon$ although ensuring feasibility, does not ensure optimality. Another aspect that can be noted is that for a penalty $\rho$ = 0.0001 the algorithm did reach the optimal solution, hinting at a possible relation between the final objective value and the penalty. To further explore this relation, the DEM model via the tradition-al ADMM is executed with the same penalties, and the model is allowed to run for 1000 iterations. The results are plotted in Fig. 6, and they show a relation that clearly demonstrates the objective value converges faster to the optimal value with lower penalties.

Table II. Sensitivity analysis of different penalty values.

| Penalty ($\rho$) | Iterations (k) | Normalized Obj. Value |
|---|---|---|
| 0.0001 | 25 | 0.99994 |
| 0.001 | 7 | 1.02076 |
| 0.01 | 5 | 1.03430 |
| 0.1 | 4 | 1.03527 |
| 1 | 3 | 1.03535 |
| 10 | 2 | 1.03535 |
| 100 | 2 | 1.03535 |

Although the results of Fig. 6 suggest that smaller penalties converge to an objective value equal to the optimal faster, this does not guarantee the solution is feasible, and even though all penalties converged to a solution quality metric lower than 0.1 after at least 25 iterations, the actual final solution quality levels of each result are not necessarily the same.

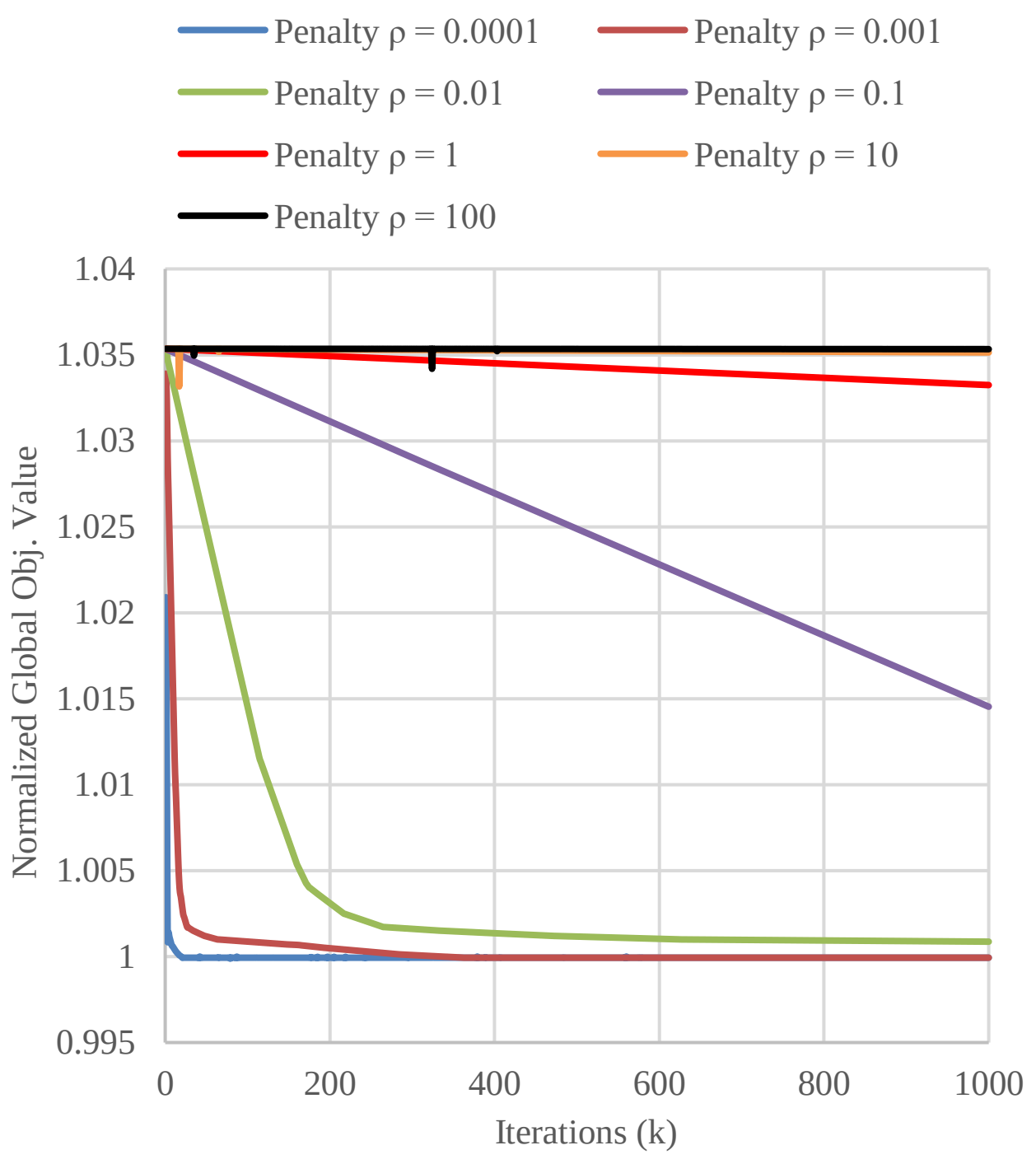

Fig. 6. Objective value convergence at different penalties.

To determine how to best identify algorithm convergence combining both metrics, the values of the solution quality metric are inspected for 1000 iterations. Closer observations of this metric indicate that eventually these values begin to oscillate between values that seem to depend on the penalty. This can be seen in Figs. 7 and 8, which show the end behavior of the solution quality in the last 300 iterations for penalties $\rho$ = 0.001 and 1, respectively, and show how the quality of the solution after 1000 iterations is improved by around a factor of 10. In addition, Table III shows the average final $\varepsilon$ value for the last 100 iterations, as well as the corresponding objective value at $k$ = 1000 for each penalty. The results indicate that with larger penalties, the solution quality is improved for the same number of iterations.

Unlike the objective value which seems to exhibit a more uniform and continuous pattern as the iterations continue, the solution quality metric decreases rapidly, but it reaches a point in which it begins to oscillate around some value that does not seem to decrease any further. Therefore, if the algorithm is programmed with the stopping criterion set to a threshold for $\varepsilon$ that the current choice of $\rho$ cannot reach, the algorithm will never stop. This can happen for example if the stopping threshold is set to 0.01 for the case when the penalty is $\rho$ = 0.001. Thus, a more robust method that can ensure the algorithm will stop for most cases would be preferred, in addition to ensuring the final solution is near the optimal and not just simply a feasible but suboptimal solution.

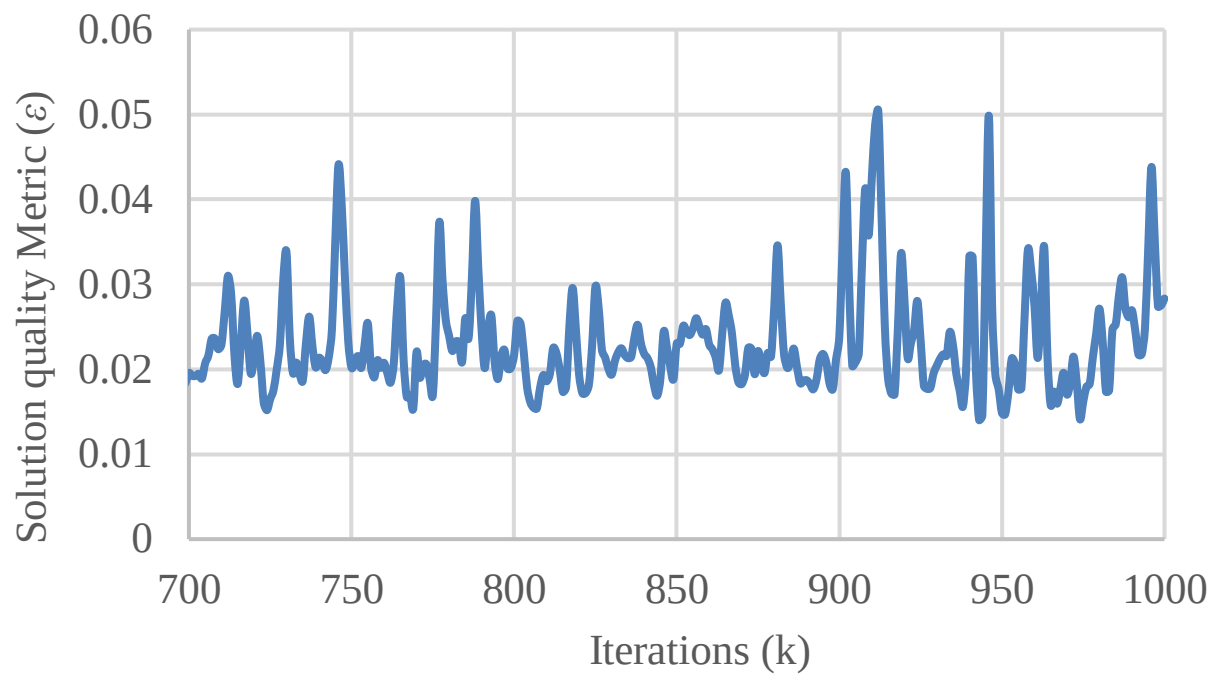

Fig. 7. Solution quality metric convergence end behavior for $\rho$ = 0.001.

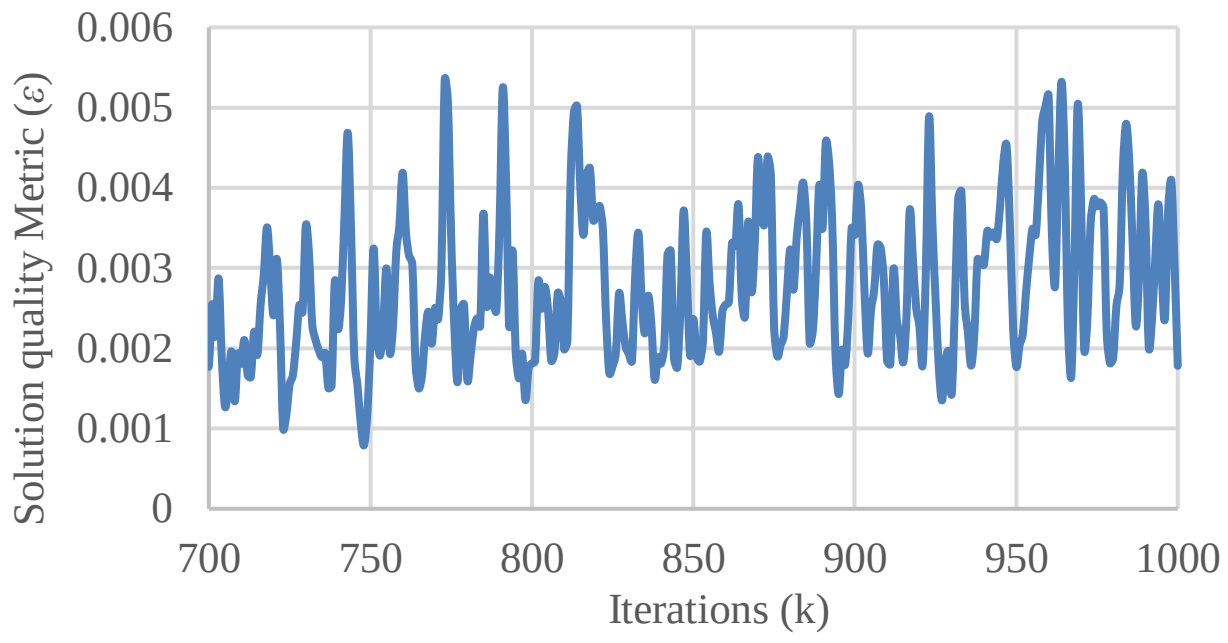

Fig. 8. Solution quality metric convergence end behavior for $\rho$ = 1.

Table III. Average final solution quality of the last 100 iterations at $k$ = 1000.

| Penalty ($\rho$) | Avg. Final Solution Quality Metric $\varepsilon$ | Final Normalized Obj. Value |
|---|---|---|
| 0.0001 | 0.08960 | 0.99994 |
| 0.001 | 0.02471 | 0.99994 |
| 0.01 | 0.03387 | 1.00087 |
| 0.1 | 0.01839 | 1.01453 |
| 1 | 0.00306 | 1.03324 |
| 10 | 0.00107 | 1.03514 |
| 100 | 0.00220 | 1.03533 |

### C. Stopping Parameters and Criteria Selection

Based on the convergence analysis of the previous subsection, a combination of the objective value and the solution quality metric as stopping criteria can be appropriate for an ADMM implementation of a decentralized networked MEM. Looking at the convergence behavior of the case with penalty $\rho$ = 0.001, the results of Fig. 6 showed that the objective value's rate of change as a function of the number of iterations decreases at around $k$ = 30, and past $k$ = 300 it seems to flatten even more. In addition, the moving average for every 100 iterations shown in Fig. 9 indicates that the solution quality metric first decreases, and then it seems to begin oscillating around the final average expected at the end of the 1000 iterations at around $k$ = 470. Based on this information it could be inferred that at this point the algorithm has already found a good solution, that is, the objective value is near the global optimal and the solution quality metric is low, near 0.03 in this case. This is the premise of the proposed DEM OB-ADMM formulation.

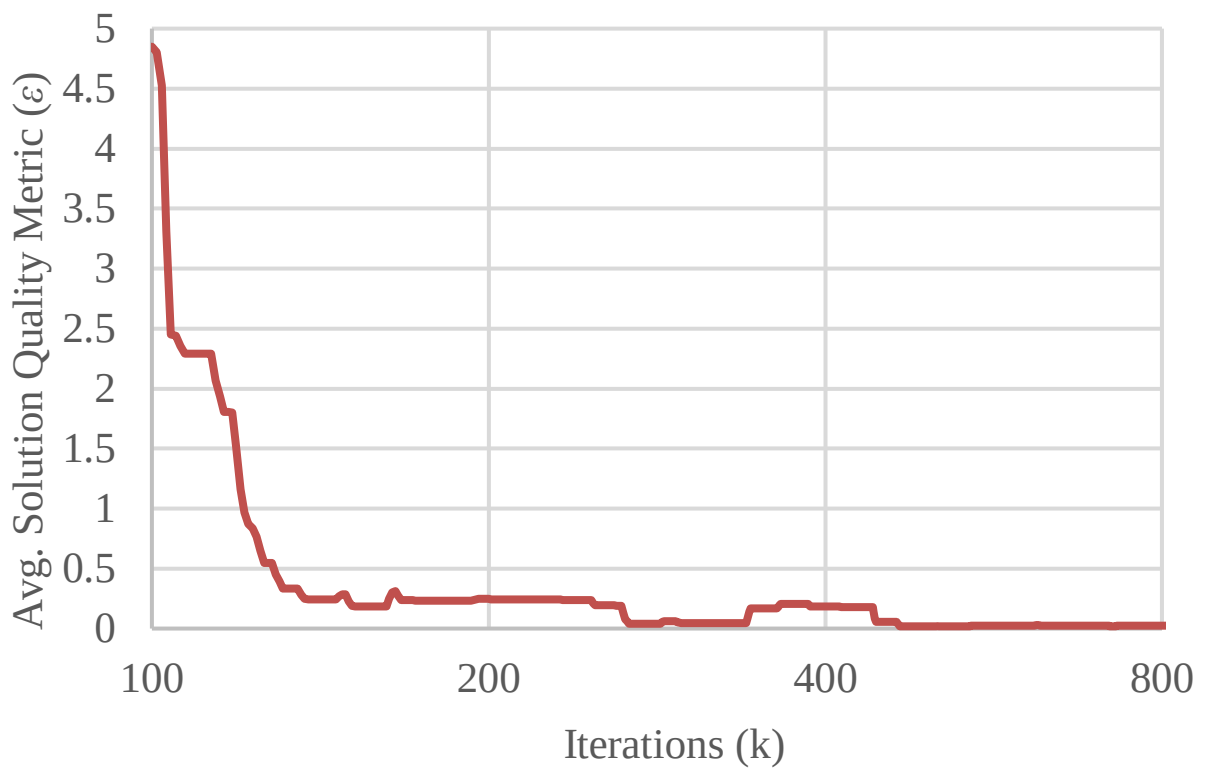

Fig. 9. Moving average of $\varepsilon$ with period 100 for $\rho$ = 0.001

The general strategy of the algorithm in Fig. 2 implements the proposed stopping criteria that combine the solution quality metric and the objective value's rate of change. The global optimal objective value of network case 1 is 3603.10 as obtained by the results of the CEM model implementation. However, in a practical scenario, the global optimal is not known, therefore a preset threshold $\beta$ for the objective value's rate of change must be established beforehand. For this case, two thresholds are tested: 0.001 and 0.01. The idea is that after a certain number of iterations, the average rate of change is inspected and once this value drops below the threshold, the solution quality metric is inspected by taking its average as well and if the current solution presents a solution quality metric value lower than this average, the algorithm will stop and collect the current solution as final. This is done so to ensure that once the rate of change of the objective value is low enough to be considered near the optimal, a solution with a quality corresponding to the valleys rather than the peaks seen in Figs. 7 and 8 is selected. This theory is tested using Fig. 2's algorithm for different penalties and the results are summarized in Table IV.

Table IV. Convergence analysis for combined stopping criteria strategy.

| Avg. Obj. Value Threshold ($\beta$) | Penalty ($\rho$) | Iterations (k) | Normalized Obj. Value | Feasibility Metric ($\varepsilon$) |
|---|---|---|---|---|
| 0.001 | 0.0001 | 680 | 0.99994 | 0.06819 |
| | 0.001 | 455 | 0.99994 | 0.01970 |
| | 0.01 | 1504 | 1.00070 | 0.00192 |
| | 0.1 | 3371 | 1.00150 | 0.03386 |
| 0.01 | 0.0001 | 344 | 0.99994 | 0.06597 |
| | 0.001 | 367 | 0.99994 | 0.02063 |
| | 0.01 | 374 | 1.00142 | 0.03583 |
| | 0.1 | 2212 | 1.00242 | 0.02397 |

For both thresholds analyzed, in terms of both objective value and solution quality, the penalty $\rho$ = 0.001 seems to be the best option for the DEM model.

### D. OB-ADMM Test Results

With the proposed OB-ADMM formulation for the DEM model already established and analyzed, more general tests are carried out using network cases 1, 2 and 3 detailed in the Appendix. The convergence results based around the objective value and solution quality metric are summarized in Table V, Table VI, and Table VII, for cases in which the penalty $\rho$ is held constant at 0.001. Furthermore, an iteration offset $k_s$ for the moving averages of 50 and 100 is tested for each case and threshold $\beta$.

These results indicate the proposed OB-ADMM algorithm effectively reaches a solution for various cases with different parameters. Moreover, the solution with the highest objective value only deviated by 0.013% from the actual optimal solution obtained with the centralized model, and the solution quality metric is well under 0.04 for all three network cases with the different hyperparameter combinations tested.

Table V. OB-ADMM test results for network case 1.

| Iteration Offset ($k_s$) | Avg. Obj. Value Change Threshold ($\beta$) | Iterations ($k$) | Normalized Obj. Value | Solution Quality Metric ($\varepsilon$) |
|---|---|---|---|---|
| 100 | 0.001 | 455 | 0.99994 | 0.01970 |
| 100 | 0.01 | 367 | 0.99994 | 0.02063 |
| 50 | 0.001 | 410 | 0.99994 | 0.02123 |
| 50 | 0.01 | 327 | 1.00002 | 0.02165 |

Table VI. OB-ADMM test results for network case 2.

| Iteration Offset ($k_s$) | Avg. Obj. Value Change Threshold ($\beta$) | Iterations ($k$) | Normalized Obj. Value | Solution Quality Metric ($\varepsilon$) |
|---|---|---|---|---|
| 100 | 0.001 | 282 | 1.00000 | 0.00000 |
| 100 | 0.01 | 206 | 1.00000 | 0.03219 |
| 50 | 0.001 | 237 | 1.00000 | 0.03599 |
| 50 | 0.01 | 187 | 1.00000 | 0.02864 |

Table VII. OB-ADMM test results for network case 3.

| Iteration Offset ($k_s$) | Avg. Obj. Value Change Threshold ($\beta$) | Iterations ($k$) | Normalized Obj. Value | Solution Quality Metric ($\varepsilon$) |
|---|---|---|---|---|
| 100 | 0.001 | 252 | 1.00000 | 0.01462 |
| 100 | 0.01 | 146 | 1.00006 | 0.03327 |
| 50 | 0.001 | 234 | 1.00000 | 0.01221 |
| 50 | 0.01 | 102 | 1.00013 | 0.02724 |

In addition, it can be noted that the use of lower $\beta$ values boosts the convergence time by reducing the number of iterations needed, but the solution quality also decreases, while a lower iteration offset $k_s$ sometimes may achieve faster convergence with similar solution quality and objective value.

The PEA from Algorithm 1 is now implemented for the DEM model, and the individual microgrid objective value results for each network case for the hyperparameters $\rho$ = 0.001, $k_s$ = 100, and $\beta$ = 0.001 with and without the PEA are shown in Table VIII and Table IX, respectively.

Table VIII. Objective value results without PEA implementation.

| Microgrid | Case 1 Obj. Value | Case 2 Obj. Value | Case 3 Obj. Value |
|---|---|---|---|
| 1 | 241.66 | 1153.79 | 499.98 |
| 2 | 3258.83 | 3222.13 | 551.88 |
| 3 | 74.46 | 106.92 | 232.93 |
| 4 | 27.92 | - | -0.83 |
| TOTAL | 3602.87 | 4482.84 | 1283.96 |

From Table IX, it can be noted that the objective value results of each microgrid after the PEA implementation to the resulting solution from the proposed DEM model are closer to those obtained with the CEM model from Table VII. This shows that with the implementation of the PEA, the final global solution of the network is guaranteed to have higher fidelity to the actual optimal solution that yields equally fair economic benefit for all participants.

Table IX. Object value results with PEA implementation.

| Microgrid | Case 1 Obj. Value | Case 2 Obj. Value | Case 3 Obj. Value |
|---|---|---|---|
| 1 | 243.61 | 1153.95 | 499.70 |
| 2 | 3258.83 | 3221.97 | 552.20 |
| 3 | 73.32 | 106.92 | 233.09 |
| 4 | 27.11 | - | -1.03 |
| TOTAL | 3602.87 | 4482.84 | 1283.96 |

## V. Conclusions

This paper proposes a novel enhanced ADMM algorithm, objective-based ADMM, for coordinating a group of interconnected networked microgrids. The proposed OB-ADMM guarantees convergence to a solution that is both high-quality and optimal. The OB-ADMM decentralized optimization model for a network of microgrids derived in this paper achieves the following combined benefits:

- Collective energy management of multiple microgrids allowing for strategic local network power exchanges as well as direct power exchanges with the main grid for all participating microgrids.
- Decentralized, privacy-preserving network operation with guaranteed high fidelity of optimization solution as compared to possible solutions obtained with an equivalent centralized optimization of the same network.
- Economically balanced network solution in which all participants are ensured to receive equal economic benefit proportional to their power import/export requirements, based on a network strategy in which local network exchanges are more economically beneficial than main grid exchanges.

The convergence and performance analyses over different test cases show that the proposed OB-ADMM formulation for the DEM of multiple microgrids is able to effectively obtain a solution with a high degree of global solution quality.

## Appendix

The parameters for the three synthetic test cases used in this paper are included in this Appendix, with all the relevant data and information relevant to the different mix of diesel generators, battery energy storage units, solar and wind power generation, and regular load demands come from various online sources [22]-[26].

The first test case corresponds to a network of four microgrids, and its parameters for generators and energy storage are given in Table X and Table XI, respectively, while the second test case corresponds to a network of three microgrids, and its parameters for generator and energy storage units are given in Table XII and Table XIII, respectively. Lastly, a third test case corresponding to another network of four microgrids but with different generator and load profiles is given, with parameters for its generators and energy storage units given in Table XIV and Table XV, respectively. The net load profiles for each corresponding network case are given in Fig. 10, 11, and 12. These correspond to a combination of load demand of end-consumers and power input of solar and wind generation for the entire 24-hour period.

Table X. Generator parameters for network case 1.

| $m$ | $g$ | $P^G_{min,m,g}$ [kW] | $P^G_{max,m,g}$ [kW] | $C^G_{m,g}$ [$/kWh] | $SU^G_{m,g}$ [$] | $NL^G_{m,g}$ [$/h] |
|---|---|---|---|---|---|---|
| 1 | 1 | 100 | 650 | 0.33 | 15.00 | 11.00 |
| | 2 | 150 | 800 | 0.28 | 13.00 | 8.70 |
| | 3 | 0 | 0 | 0 | 0 | 0 |
| 2 | 1 | 100 | 650 | 0.33 | 15.00 | 11.00 |
| | 2 | 150 | 800 | 0.28 | 13.00 | 8.70 |
| | 3 | 260 | 940 | 0.23 | 10.35 | 7.40 |
| 3 | 1 | 250 | 900 | 0.26 | 11.85 | 8.45 |
| | 2 | 0 | 0 | 0 | 0 | 0 |
| | 3 | 0 | 0 | 0 | 0 | 0 |
| 4 | NO GENERATORS | | | | | |

Table XI. Energy storage parameters for network case 1.

| $m$ | $b$ | $P^{ES}_{lim,m,b}$ [kW] | $EL^{ES}_{min,m,b}$ [kWh] | $EL^{ES}_{max,m,b}$ [kWh] | $\eta^{ESc}_{m,b}$ [-] | $\eta^{ESd}_{m,b}$ [-] |
|---|---|---|---|---|---|---|
| 1 | 1 | 1000 | 500 | 5000 | 0.96 | 0.98 |
| | 2 | 500 | 350 | 3500 | 0.96 | 0.98 |
| 2 | 1 | 800 | 420 | 4200 | 0.96 | 0.98 |
| | 2 | 400 | 240 | 2400 | 0.96 | 0.98 |
| 3 | 1 | 1000 | 500 | 5000 | 0.96 | 0.98 |
| | 2 | 500 | 350 | 3500 | 0.96 | 0.98 |
| 4 | 1 | 1000 | 500 | 5000 | 0.96 | 0.98 |
| | 2 | 500 | 350 | 3500 | 0.96 | 0.98 |

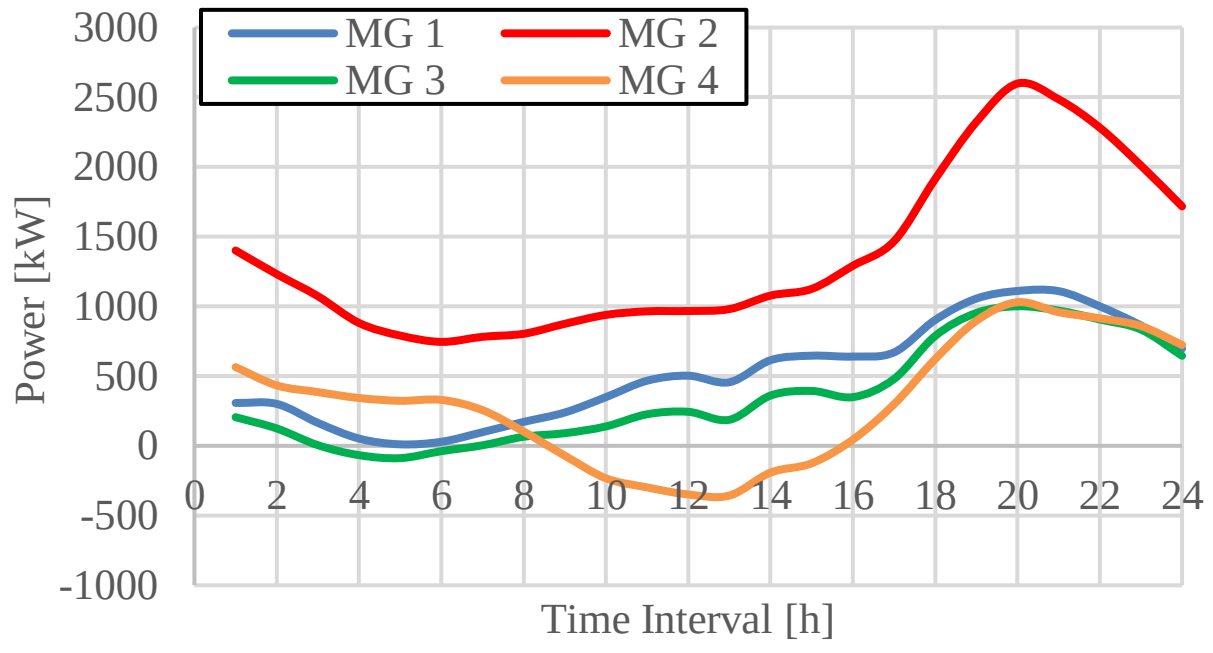

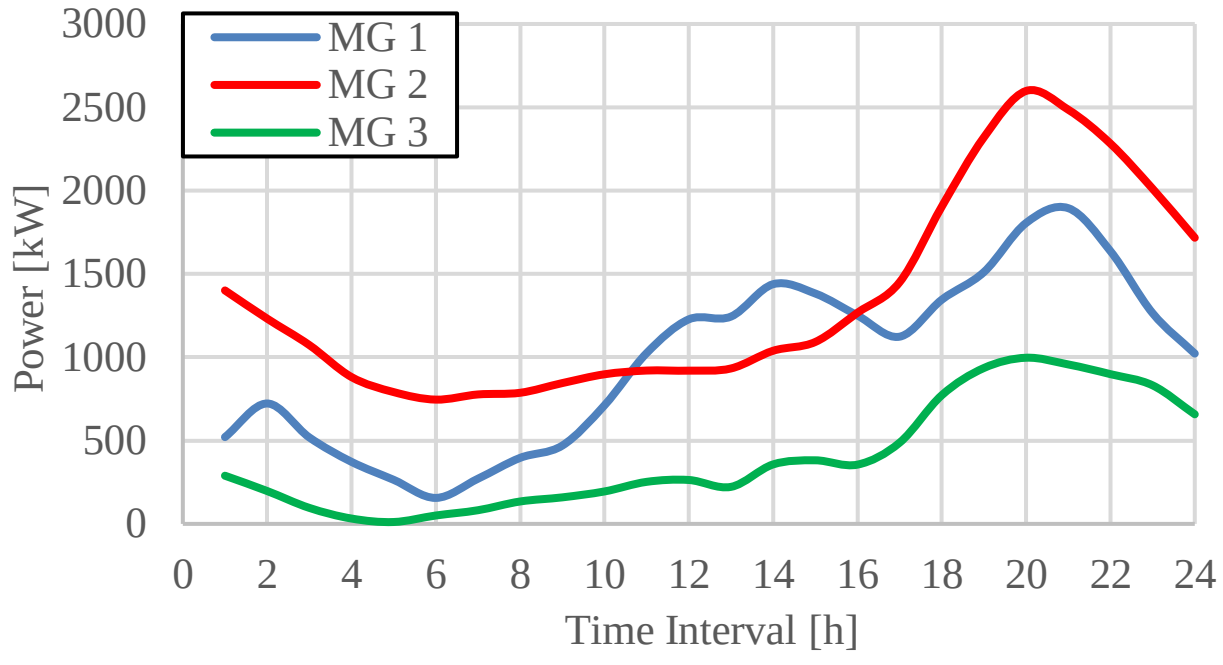

Fig. 11. Net load profiles for network case 2.

Table XII. Generator parameters for network case 2.

| $m$ | $g$ | $P^G_{min,m,g}$ [kW] | $P^G_{max,m,g}$ [kW] | $C^G_{m,g}$ [$/kWh] | $SU^G_{m,g}$ [$] | $NL^G_{m,g}$ [$/h] |
|---|---|---|---|---|---|---|
| 1 | 1 | 100 | 650 | 0.33 | 15.00 | 11.00 |
| | 2 | 150 | 800 | 0.28 | 13.00 | 8.70 |
| 2 | 1 | 250 | 900 | 0.26 | 11.85 | 8.45 |
| | 2 | 260 | 940 | 0.23 | 10.35 | 7.40 |
| 3 | 1 | 200 | 825 | 0.27 | 12.00 | 8.60 |
| | 2 | 0 | 0 | 0 | 0 | 0 |

Table XIII. Energy storage parameters for network case 2.

| $m$ | $b$ | $P^{ES}_{lim,m,b}$ [kW] | $EL^{ES}_{min,m,b}$ [kWh] | $EL^{ES}_{max,m,b}$ [kWh] | $\eta^{ESc}_{m,b}$ [-] | $\eta^{ESd}_{m,b}$ [-] |
|---|---|---|---|---|---|---|
| 1 | 1 | 500 | 3500 | 350000 | 0.96 | 0.98 |
| | 2 | 500 | 350 | 3500 | 0.96 | 0.98 |
| 2 | 1 | 800 | 420 | 4200 | 0.96 | 0.98 |
| | 2 | 400 | 240 | 2400 | 0.96 | 0.98 |
| 3 | 1 | 1000 | 500 | 5000 | 0.96 | 0.98 |
| | 2 | 400 | 240 | 2400 | 0.96 | 0.98 |

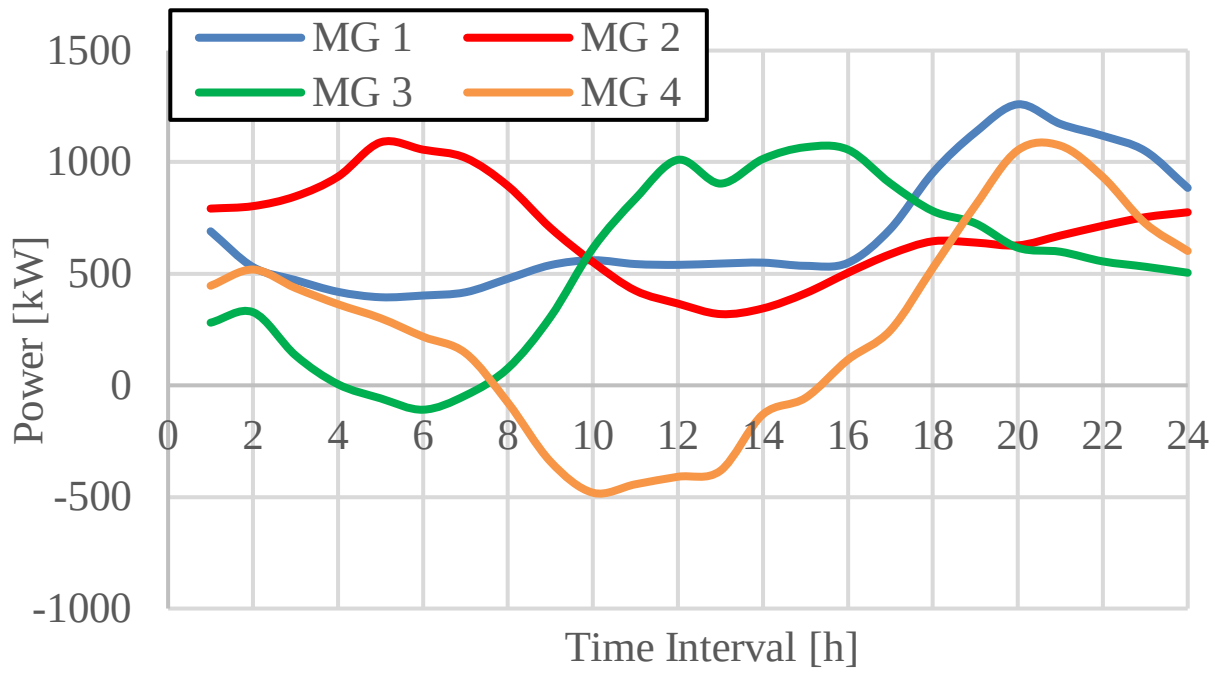

Fig. 12. Net load profiles for network case 3.

Table XIV. Generator parameters for network case 3.

| $m$ | $g$ | $P^{G}_{min,m,g}$ [kW] | $P^{G}_{max,m,g}$ [kW] | $C^{G}_{m,g}$ [$/kWh] | $SU^{G}_{m,g}$ [$] | $NL^{G}_{m,g}$ [$/h] |
|---|---|---|---|---|---|---|
| 1 | 1 | 150 | 800 | 0.28 | 13.00 | 8.70 |
| | 2 | 0 | 0 | 0 | 0 | 0 |
| 2 | 1 | 260 | 940 | 0.23 | 10.35 | 7.40 |
| | 2 | 0 | 0 | 0 | 0 | 0 |
| 3 | 1 | 100 | 650 | 0.33 | 15.00 | 11.85 |
| | 2 | 250 | 900 | 0.26 | 11.85 | 8.45 |
| 4 | NO GENERATORS | | | | | |

Table XV. Energy storage parameters for network case 3.

| $m$ | $b$ | $P^{ES}_{lim,m,b}$ [kW] | $EL^{ES}_{min,m,b}$ [kWh] | $EL^{ES}_{max,m,b}$ [kWh] | $\eta^{ESc}_{m,b}$ [-] | $\eta^{ESd}_{m,b}$ [-] |
|---|---|---|---|---|---|---|
| 1 | 1 | 800 | 420 | 4200 | 0.96 | 0.98 |
| | 2 | 500 | 350 | 3500 | 0.96 | 0.98 |
| 2 | 1 | 500 | 350 | 3500 | 0.96 | 0.98 |
| | 2 | 400 | 240 | 2400 | 0.96 | 0.98 |
| 3 | 1 | 800 | 420 | 4200 | 0.96 | 0.98 |
| | 2 | 400 | 240 | 2400 | 0.96 | 0.98 |
| 4 | 1 | 1000 | 500 | 5000 | 0.96 | 0.98 |
| | 2 | 500 | 350 | 3500 | 0.96 | 0.98 |